\documentclass[aps,pra,twocolumn,superscriptaddress]{revtex4-1}

\usepackage[english]{babel}
\usepackage{braket}
\usepackage{mathtools}
\usepackage{amssymb,bm}
\usepackage{tabularx, booktabs}
\usepackage{tikz,pgfplots}
\usepackage{url}
\usepackage{graphicx}
\usepackage{epsfig}
\usepackage{color}
\usepackage{array}
\usepackage{csquotes}
\usepackage{xpatch}
\usepackage{xspace}
\usepackage[normalem]{ulem} 
\usepackage{physics}
\usepackage{appendix}
\usepackage{bbold} 
\usepackage[colorlinks=true, linkcolor=blue, citecolor=dred]{hyperref}
\hypersetup{
    colorlinks=true,
    citecolor=gray,
    filecolor=blue,
    linkcolor=blue,
    urlcolor=red
}

\definecolor{ICGMmarine}{rgb}{0.168, 0.168, 0.525}
\definecolor{ICGMblue}{rgb}{0, 0.549, 0.714}
\definecolor{ICGMorange}{rgb}{0.968, 0.647, 0}
\definecolor{ICGMyellow}{rgb}{1, 0.804, 0}
\definecolor{GPT4}{RGB}{16,163,127}

\definecolor{dgreen}{rgb}{0,.5,0}
\definecolor{dred}{rgb}{.7,.0,.0}

\newcommand{\Eq}[1]{Eq.~(\ref{#1})}

\newcommand{\Eqss}[2]{Eqs.~(\ref{#1}) and~(\ref{#2})}

\newcommand{\Fig}[1]{Fig.~(\ref{#1})}

\newcommand{\Sec}[1]{Sec.~\ref{#1}}

\newcommand{\etal}{{\it et al.}}

\newcommand{\be}{\begin{eqnarray}}
\newcommand{\ee}{\end{eqnarray}}

\usepackage{xr}

\begin{document}
\title{
Unitary transformations within density matrix embedding approaches: A novel perspective on the self-consistent scheme for electronic structure calculation
}
\author{Quentin Mar\'ecat}
\affiliation{ICGM, Université de Montpellier, CNRS, ENSCM, 34000 Montpellier (France)}
\author{Benjamin Lasorne}
\affiliation{ICGM, Université de Montpellier, CNRS, ENSCM, 34000 Montpellier (France)}
\author{Emmanuel Fromager}
\affiliation{Laboratoire de Chimie Quantique, Institut de Chimie, CNRS, Université de Strasbourg, 4 rue Blaise Pascal, 67000 Strasbourg (France)}
\author{Matthieu Sauban\`ere}
\email{matthieu.saubanere@cnrs.fr}
\affiliation{ICGM, Université de Montpellier, CNRS, ENSCM, 34000 Montpellier (France)}
\begin{abstract}
In this work, we introduce an original self-consistent scheme based on the one-body reduced density matrix ($\gamma$) formalism. A significant feature of this methodology is the utilization of an optimal unitary transformation of the Hamiltonian, determined through a self-consistently determined, unitary reflection $\mathbf{R}[\gamma]$. This enables the extraction of all reduced properties of the system from a smaller, accurately solved embedding cluster, and to systematically reconstruct the reduced density matrix of the system. This process ensures that both extended and embedded systems satisfy the local virial-like relation, providing quantitative insight into the correspondence between the fragment in the extended system and its embedded analogue.
The performance and convergence of the method, as well as the N-representability of the resulting correlated density matrix, are evaluated and discussed within the context of the one-dimensional Hubbard model, which provides exact results for a comprehensive comparison.
\end{abstract}
\maketitle

\section{Introduction}

Given the complexity associated with the resolution of the Schr$\ddot{\rm o}$dinger equation, Divide and Conquer (DaC) strategies have gained significant attention in the past decade to approch the electronic structure of strongly interacting electron systems~\cite{kotliar_cellular_2001,lupo_maximally_2021,nguyen_lan_rigorous_2016,zheng_self-consistent_1993,ma_quantum_2021}.  
More specifically, these strategies entail in dividing a full interacting problem into smaller, {\it i.e.} tractable pieces that can be individually treated before reconstructing the complete solution to the original extended problem.
The development of such algorithms hinges on two crucial aspects. First, there is a need for methods to construct various tractable pieces or reduced effective systems for which properties can be computed. Second, there is a requirement for a methodology to reconstruct the properties of the original extended system based on the collected properties of its constituent pieces.

In this context, various quantum embedding approaches have received particular attention over the past decade to tackle the first aspect by partitioning the extended many-body system into smaller pieces. Quantum embedding efficiently allows for the study of properties of a subset of atoms within a larger system. This is achieved by representing this subset with a minimal number of effective degrees of freedom that capture all the remaining effects of the entire system~\cite{sun_quantum_2016,wasserman_quantum_2020}.
Embedding can take various forms depending on the theoretical framework used to describe the full extended systems. These forms include local or non-local effective potentials acting on original or additional orbitals within the density ~\cite{gordon_theory_1972,gordon_density_2017,govind_electronic-structure_1999,senjean_local_2017,senjean_site-occupation_2018,senjean_projected_2019}, the reduced density-matrix~\cite{schade2018adaptive,sekaran_householder_2021,sekaran_local_2022,yalouz_quantum_2022,lanata2023derivation}, the wave function~\cite{lacombe_embedding_2020,nusspickel2022systematic,marecat_recursive_2022}, or the Green's function frameworks~\cite{muller-hartmann_correlated_1988,georges_dynamical_1996,georges_hubbard_1992,senechal_spectral_2000,mazouin_site-occupation_2019,weng_embedded_2022,ayral_dynamical_2017,zgid_dynamical_2011}.

Among the different formalisms, the one-particle reduced density matrix (1-RDM) formalism looks appealing since it contains all one-body non-local information of the system contrary to the density, and can easily be constructed because the N-representable conditions are known~\cite{coleman1963structure}.
In addition, 1-RDM have been shown to be a pertinent quantity to establish embedding recipies. 
In particular, the density matrix embedding theory (DMET)~\cite{knizia_density_2012} has been recently highlighted due to its ability to capture local correlation in molecules and periodic systems~\cite{wouters_practical_2016,bulik_density_2014,cui_efficient_2020,zheng_cluster_2017,ye_bootstrap_2019,ricke_performance_2017}. Within the DMET method, the extended system is divided into several {\it fragments}. Each fragment is separately embedded with few additional orbitals called {\it bath} such that the solution of the fragment+bath {\it cluster} is affordable using a full configuration interaction approach. 
Instead of explicitly reconstructing the solution of the extended system from the local properties of the fragments, DMET achieves self-consistency by searching an effective low-level auxiliary system in which the local 1-RDMs match those of the embedding cluster.
This approach has been shown to provide an efficient, accurate and affordable way for studying the electronic structure of strongly correlated materials~\cite{pham_periodic_2020,cui_ground-state_2020,mitra_periodic_2022}.
Recently, Cances et al. have carried out rigorous investigation of DMET into the mathematical aspects, as detailed in their study ~\cite{cances2023mathematical}. In addition, extensions to DMET have been proposed for the study of strongly correlated systems, which allows for the treatment of open-shell systems~\cite{mitra_excited_2021} and the calculation of excited states with the energy-weighted RDM~\cite{fertitta_energy-weighted_2019}, for example.

In this paper, we present an alternative self-consistent DaC algorithm to access ground state properties of correlated systems that is fully based on the 1-RDM. Concerning the division of the full system we follow the recent work of Sekaran \etal ~\cite{sekaran_householder_2021,sekaran_local_2022,yalouz_quantum_2022,sekaran_unified_2023} that provide a systematic and optimal construction of embedding clusters by means of a unitary transformation in the one-electron Hilbert space, which can be interpreted geometrically as a reflection and is defined as a functional of the 1-RDM. The exact wave function of the cluster and reduced quantities such as the local 1- and 2-RDMs can be computed for instance using the Lanczos algorithm. Importantly, we propose here to achieve the ``conquest'' by introducing a reconstruction scheme for the full-system 1-RDM from all cluster 1-RDMs. More precisely, we define self-consistent conditions for obtaining iterative DaC algorithms. Note that the reconstruction protocol allows to obtain non-idempotent 1-RDM for the extended system, {\it i.e.} that are not associated to a single Slater determinant. As a proof of concept, results are presented and compared with the standard DMET approach in the context of the one-dimensional Hubbard model, for which exact results are available. The extension to ab-initio Hamiltonians is straightforward and discussed in light of the presented and encouraging results hereafter.

\section{Theory}\label{theory}
  
\begin{figure}
    \centering
    \resizebox{\columnwidth}{!}{
	\includegraphics[scale=0.20]{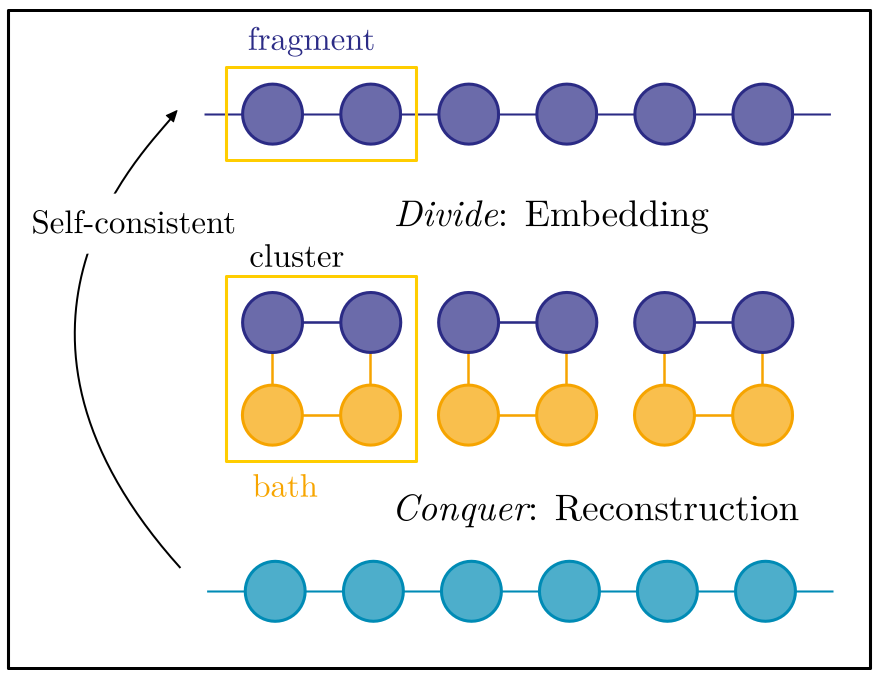}
    }
    \caption{Schematic representation of the divide and conquer strategy.}
    \label{fig:Div-Conq}
  \end{figure}

In this section we propose a self-consistent Divide-and-Conquer like algorithm based on a reflection, functional of the 1-RDM introduced previously. As shown schematically in \Fig{fig:Div-Conq} the process is divided into two steps. At the \textit{Divide} step, the full problem is split into many fragments. Following an embedding approach, each fragment is complemented by additional bath degrees of freedom that optimally mimic the original environment. The fragment+bath (cluster) has to be small enough such that the calculation of their exact properties is affordable. The \textit{Conquer} step consists in using the computed properties of all clusters to reconstruct the full-size problem. 
  For the sake of clarity we focus on the periodic, paramagnetic 1D Hubbard Hamiltonian~\cite{hubbard_electron_1963} given by
  \begin{eqnarray}\label{eq:hubbard}
    \hat{H} &=& \hat{h}^0 + \hat{W}, \\
     \hat{h}^0 &=& -t\sum_{\langle i,j \rangle\sigma}\left( \hat{c}^\dagger_{i\sigma}\hat{c}_{j\sigma} +  \hat{c}^\dagger_{j\sigma}\hat{c}_{i\sigma}\right), \label{eq:kin} \\
     \hat{W} &=& U\sum_i \hat{n}_{i\uparrow}\hat{n}_{i\downarrow,}
  \end{eqnarray}
  where $\hat{c}^\dagger_{i\sigma}$ ($\hat{c}_{i\sigma}$) corresponds to creation (annihilation)  of an electron of spin $\sigma$ in the atomic orbital $i$, respectively and $\hat{n}_{i\sigma}$ is the spin-density operator equal to $\hat{c}^\dagger_{i\sigma}\hat{c}_{i\sigma}$. The first (one-body) operator in \Eq{eq:hubbard} corresponds to the kinetic operator with $-t$ being the hopping integral and the subscripts $<i,j>$ refer to pairs of nearest neighbour (NN) orbitals, while the second (two-body) operator accounts for the electron-electron interaction, where $U$ refers to the on-site Coulomb integral. 
    \subsection{Unitary reflection functional of the reduced density-matrix} \label{sec:emb}
In this subsection we recall the embedding strategy based on the block-Householder unitary transformation of the one-body density matrix, introduced (and discussed in detail) in the following Refs.~\cite{yalouz_quantum_2022,sekaran_unified_2023}.
Let us consider a  {\it fragment} containing  $N_i$ impurities, $1\leq  N_i \leq N_s/2$ with $N_s$ the number of spatial orbital in the system equal to the number of sites in the Hubbard lattice, each spatial orbital containing two spin channels $\sigma$. For each fragment and spin channels $\sigma$, we introduce a generic construction of a unitary reflection $\mathbf{R}^\sigma=\mathbf{R}^\sigma[\gamma^\sigma]$, functional of the spin one-body reduced density matrix $\gamma^\sigma$ as \\
\begin{eqnarray}\label{eq:symP}
 \mathbf{R}^\sigma = \mathbb{1} - 2\mathbf{V}^\sigma(\mathbf{V}^{\sigma T}\mathbf{V}^\sigma)^{-1}\mathbf{V}^{\sigma T}.
\end{eqnarray}
$\mathbf{R}^\sigma$ is, by construction, an involutory matrix ($\mathbf{R}^{\sigma 2} = \mathbb{1}$). Furthermore, it is real symmetric ($\mathbf{R}^{\sigma^T}=\mathbf{R}^{\sigma}$). Consequently, it is an orthogonal matrix and represents an isometry, and is sometimes denoted as a normal involution.
More specifically, the mathematical nature of the transformation $\mathbf{R}^\sigma$ depends on the value of its determinant, which is equal to $(-1)^{N_i}$. 
When $N_i$ is an even number, the determinant is $1$, and the transformation constitutes a rotation. Conversely, when $N_i$ is an odd number, the determinant is $-1$, indicating an improper rotation. The exceptional case of $N_i=1$ corresponds to a reflection.
In all cases, the eigenvectors associated with the eigenvalues $1$ remain invariant under the transformation, while those connected to the eigenvalues $-1$ are flipped, and can be interpreted geometrically as a reflection.
In that sense, the involution $\mathbf{R}^\sigma$ is denoted in the following as a reflection for clarity. The one-particle density matrix $\gamma^\sigma_{ij} = \bra{\Psi}\hat{c}^\dagger_{i\sigma} \hat{c}_{j\sigma}\ket{\Psi}$ is assumed to be known, where $\ket{\Psi}$ denotes the many-body ground state wave-function with the two spin channels assumed to be uncorrelated. More precisely and as detailed in Appendix~\ref{app:reflection}, $\mathbf{V}^\sigma = \mathbf{V}^\sigma[\gamma^\sigma] \in \mathbb{R}_{N_sN_i}$ must be constructed so that $\tilde{\gamma}^\sigma = \mathbf{R}^\sigma\gamma^\sigma\mathbf{R}^{\sigma}$ entangles the fragment optimally with few bath orbitals. 
For the sake of clarity, quantities that have been transformed using the unitary reflection $\mathbf{R}^\sigma$ are expressed in the $\mathbf{R}$-representation (or basis), and are highlighted by a tilde $\tilde{\,\,}$ symbol. 
Importantly, the number of orbitals within the bath corresponds to the number of orbitals in the fragment,  as in DMET~\cite{knizia_density_2012,sekaran_unified_2023}.
Moreover, orbitals in the fragment are preserved in the $\mathbf{R}$-representation such that\\
\begin{eqnarray}\label{eq:impkeep}
 \tilde{\gamma}^\sigma_{ij} = \gamma^\sigma_{ij} \quad i,j\in \text{fragment},
\end{eqnarray}
and are fully disentangled from the environment at the one-body level \\
\begin{eqnarray}\label{eq:dis}
 \tilde{\gamma}^\sigma_{ij} = 0 \quad i\in \text{fragment, }j \in \text{environment},
\end{eqnarray}
where the environment refers here to the environment of the embedding cluster.
Interestingly, advantageous properties appear in the non interacting limit, when the ground state $| \Psi \rangle$ is reduced to a single Slater determinant 
$| \phi \rangle = \prod_{\mu \sigma}c^\dagger_{\mu \sigma}| 0 \rangle $, where $\mu$ belongs to all occupied orbitals solutions of the one-body hamiltonian $h^0$ in \Eq{eq:kin} (Bloch states) and $| 0 \rangle$ the vacuum state.
This single Slater determinant is associated to an idempotent density-matrix ($\gamma^\sigma = \gamma^{\sigma2}$). 
In this case, Sekaran \etal{} demonstrated (Appendix B of~\cite{sekaran_householder_2021}) that the cluster {\it (i)} is fully disentangled from the environment, {\it i.e.}
\begin{equation}
  \label{eq:condMF0}
\tilde{\gamma}^\sigma_{ij} = 0 \quad i\in \text{cluster, } j \in \text{environment}
\end{equation}
and {\it (ii)} 
the number of electrons with spin $\sigma$ is an integer equal to $N_i$,
\begin{equation}
  \label{eq:condMF1}
  {\rm tr}\left[\tilde{\gamma}^\sigma|_c\right] = N_i,
\end{equation}
where $|_c$ refers to all orbitals belonging to the cluster. Within a wave-function formalism, properties~(\ref{eq:condMF0}) and~(\ref{eq:condMF1}) correspond to a  perfect factorization of the wave function
\begin{equation}
  \label{eq:condMF2}
  | \tilde{\phi} \rangle = \mathbf{R}| \phi \rangle = \hat{\mathcal{A}}|\tilde{\phi}^c\rangle |\tilde{\phi}^e\rangle.
\end{equation}
where $|\tilde{\phi}^c\rangle$ ($|\tilde{\phi}^e\rangle$) describes the cluster (environment), respectively, and the operator $\hat{\mathcal{A}}$ is the antisymmetrizing operator.
%
$ \mathbf{R}| \phi \rangle$ is a shortcut notation, such that the Slater determinant is expressed in the $\mathbf{R}$-representation using the following equation,
\begin{eqnarray} \label{eq:transfo1}
   \hat{c}_{i\sigma}^{\dagger} = \sum_{k}\mathbf{R}^{\sigma}_{ik}\tilde{c}_{k\sigma}^{\dagger}, &\quad \hat{c}_{i\sigma} = \sum_{k}\mathbf{R}^{\sigma}_{ik}\tilde{c}_{k\sigma}
\end{eqnarray}
 where $\tilde{c}_{k\sigma}^{\dagger}$ ($\tilde{c}_{k\sigma}$) stands for the creation (annihilation) operator of an electron in the $k$ orbital with spin $\sigma$ expressed in the $\mathbf{R}$-representation, respectively.
%
It follows, at the mean-field level, that all properties of the cluster can be exactly extracted from $|\tilde{\phi}^c\rangle$, including $\tilde{\gamma}^{\sigma c} = \langle \tilde{\phi}^c | \hat{\gamma}^\sigma|\tilde{\phi}^c\rangle  = \mathbf{R}^\sigma \gamma^\sigma \mathbf{R}^\sigma|_c$. 
Additionally, $|\tilde{\phi}^c\rangle$ corresponds to the ground state of the Hamiltonian $\tilde{h}^c$ associated to the projection onto the cluster of the non interacting Hamiltonian $\hat{h}^0$ expressed in the $\mathbf{R}$-representation.
Otherwise, when the 1-RDM deviates from idempotency ({\it i.e.} at least one eigenvalue of $\gamma$, $0 \le \eta \le 1$, strictly differs from 0 or 1),  properties~(\ref{eq:condMF0},~\ref{eq:condMF1},~\ref{eq:condMF2}) are not fulfilled anymore ~\cite{sekaran_householder_2021}. In other words, $\tilde{\gamma}^\sigma=\mathbf{R}^\sigma \gamma^\sigma \mathbf{R}^\sigma$ is not  block-diagonal. 
In the following we present approximations leading to a projection onto the cluster of the Hamiltonian, unlocking the use of the correlated, {\it i.e.} non-idempotent density matrix as a fundamental variable in an original self-consistent embedding scheme based on a divide and conquer like algorithm.
\subsection{Divide and conquer algorithm} \label{sec:sces}
For each fragment, we generalize properties~(\ref{eq:condMF0},~\ref{eq:condMF1},~\ref{eq:condMF2}) valid only for Slater determinants $\ket{\phi}$ by proposing the following approximation for the many-body ground-state wave function $|\Psi \rangle$ in the $\mathbf{R}$-representation
\begin{equation}
  \label{eq:DAC_eq}
  | \tilde{\Psi} \rangle = \mathbf{R}|\Psi \rangle \simeq \hat{\mathcal{A}} |\bar{\Psi}^c\rangle |\bar{\Psi}^e\rangle.
\end{equation}
The approximate 1-RDM functional factorization \Eq{eq:DAC_eq} is the central equation of the proposed \textit{Divide and Conquer} algorithm, that consists in iteratively obtaining $\mathbf{R}^\sigma[\gamma^\sigma]$ and $|\bar{\Psi}^c\rangle$.
In the following, the wave function denoted $\ket{\bar{\Psi}^e}$ will not be required, and therefore, will not be calculated.
 As detailed in the previous section, the reflection $\mathbf{R}^\sigma$ is defined as a functional of the 1-RDM $\gamma^\sigma$ associated to the ground state $|\Psi \rangle$. 
 Determining $|\bar{\Psi}^c\rangle$ and all local properties of the fragments is thus the main objective of what follows.
 To that aim, we transform the Hamiltonian $\hat{H}$ in the $\mathbf{R}$-representation ($\tilde{H}$) using \Eq{eq:transfo1}.
 In addition, $\tilde{H}$ can be further separated into $\bar{H}^c$ ($\bar{H}^e$) that contains terms involving only orbitals belonging to the cluster (environment), respectively, and the  cluster-environment interactions $\tilde{H}^{ce}$, {\it i.e.} $\tilde{H} = \bar{H}^c + \bar{H}^e + \tilde{H}^{ce}$, and $\bar{H}^c_{ij} = \hat{H}_{ij}$ for all $i,j$ in the fragment.
Since $\mathbf{R}^\sigma$ fully disentangles, at the one-body level, the fragment from the environment, where interactions remain solely between bath and environment orbitals, we propose to neglect $\tilde{H}^{ce}$ such that
\begin{eqnarray} \label{eq:app_dec}
 \tilde{H} \sim \bar{H} = \bar{H}^c + \bar{H}^e.  
\end{eqnarray}
It follows that the ground state of $\bar{H}$ can be factorized as $| \bar{\Psi} \rangle = \hat{\mathcal{A}}|\bar{\Psi}^c\rangle |\bar{\Psi}^e\rangle$ with $|\bar{\Psi}^c\rangle$ ($|\bar{\Psi}^e\rangle$) corresponding to the ground state of the Hamiltonian $\bar{H}^c$ ($\bar{H}^e$), respectively. 
$\mathbf{R}^\sigma$ being defined through $\gamma^\sigma$,  $\bar{H}^c$ consist also explicitly  in a functional of  $\gamma^\sigma$, {\it i.e.} $\bar{H}^c = \bar{H}^c[\gamma]$.
Unlike standard embedding approaches, such as the original DMET ~\cite{knizia_density_2012}, the two-body interactions naturally emerge in the bath orbitals due to the change in representation using \Eq{eq:transfo1}.
In addition, part of the neglected cluster-environment terms in $\tilde{H}^{ce}$ can be considered at the mean-field level in the cluster Hamiltonian through the following term
\begin{eqnarray}\label{eq:mf_ce}
\bar{H}^{\rm env}_{\rm MF}[\gamma^\sigma] = \sum_{i,j\in \rm cluster}\tilde{c}^\dagger_{i\sigma}\tilde{c}_{j\sigma}\sum_{(k,l)}\tilde{U}_{ijkl}\tilde{\gamma}^{\bar{\sigma}}_{kl}+ \rm h.c.
\end{eqnarray}
where the notation $(k,l)$ refers to pairs of orbitals in the $\mathbf{R}$-representation such that at least $k$ or $l$ belongs to the environment and  $\tilde{U}_{ijkl}=U\sum_m\mathbf{R}^{\sigma}_{im}\mathbf{R}^{\sigma}_{jm}\mathbf{R}^{\bar{\sigma}}_{km}\mathbf{R}^{\bar{\sigma}}_{lm}$. 
The mean-field treatment considers the average effects in the cluster of electrons belonging to the environment.
 Note that for the sake of clarity, all quantities related to the cluster after projection are highlighted by a bar $\bar{\;\;}$ symbol.
 As Eq.~(\ref{eq:condMF1}) is not satisfied for non idempotent matrices, the number of electron per spin $\sigma$ in the cluster becomes fractional. For practical reasons and simplicity, in this contribution we work with closed clusters, for which the number of electrons per spin $\sigma$ is fixed to be the nearest integer $N_e^{\sigma c}$ of tr$[\tilde{\gamma}^{\sigma c}]$. This consists in an approximation assuming the charge leak  between the bath and the environment being weak.
Note that, beyond the scope of this work, one could work with fractional numbers of electrons in the (open) cluster by considering an ensemble density-matrix matrix, {\it i.e.}  $\bar{\gamma}_{ij}^{\sigma c} = \omega\langle \bar{\Psi}_{N_e^{\sigma 1}}^c | \tilde{c}_{i\sigma}^{\dagger}\tilde{c}_{j\sigma} |\bar{\Psi}^c_{N_e^{\sigma 1}}\rangle + (1- \omega)\langle \bar{\Psi}_{N_e^{\sigma 2}}^c | \tilde{c}_{i\sigma}^{\dagger}\tilde{c}_{j\sigma} |\bar{\Psi}_{N_e^{\sigma 2}}^c\rangle$, with $N_e^{\sigma 1}$ and  $N_e^{\sigma 2}$ integers such that $N_e^{\sigma 1} \leq {\rm tr}[\tilde{\gamma}^{\sigma c}]\leq N_e^{\sigma 2}$ and $\omega$ the weight such that ${\rm tr}[\bar{\gamma}^{\sigma c}]= {\rm tr}[\tilde{\gamma}^{\sigma c}]$. 
  Importantly, due to the approximation in \Eq{eq:app_dec}, the number of electrons in the fragment $F$ might be impacted, {\it i.e.} tr$[\bar{\gamma}^{\sigma c}|_F] = \sum_{i\in F}\langle \bar{\Psi}^c | \tilde{c}_{i\sigma}^{\dagger}\tilde{c}_{i\sigma} |\bar{\Psi}^c\rangle \neq $tr$[ \tilde{\gamma}^\sigma|_F]$ and an additional chemical-like potential $\mu_{\rm emb}$ is added over bath orbitals to overcome this issue.
Interestingly, at half-filling, the chemical-like potential $\mu_{\rm emb}$ is null,  which implies that the density per spin of each orbital belonging to the cluster fragment is strictly equal to $0.5$ for the paramagnetic 1D Hubbard model.  In this case, the effective potential $\bar{H}^{\rm env}_{\rm MF}$ is purely local and ensures that the occupation of the embedded impurities matches the one in the lattice.
All together, we obtain the following  effective cluster Hamiltonian 
  \begin{equation}
    \label{eq:cluster_Ham}
    \bar{H}^c_{\rm eff}[\gamma^\sigma] =  \bar{H}^c[\gamma^\sigma] + \bar{H}^{\rm env}_{\rm MF}[\gamma^\sigma] + \mu_{\rm emb}\sum_{i \in {\rm bath}} \tilde{n}_{i\sigma}
  \end{equation}
 with $\tilde{n}_{i\sigma}=\tilde{c}_{i\sigma}^{\dagger}\tilde{c}_{i\sigma}$, from which $|\bar{\Psi}^c\rangle$ can be explicitly computed as the ground state in the $N_e^{\sigma c}$ electron Hilbert subspace using standard numerical diagonalization methods.
 
Thus far, the main approximations made pertain to the approximate decomposition of the wave function $\ket{\Psi}$ in the $\mathbf{R}$-representation \Eq{eq:DAC_eq} together with the rounding up of electron number to work with a closed cluster for practical reasons. This closure naturally demands the introduction of an effective chemical-like potential $\mu_{\rm emb}$ to correct the impurity density of the fragments within the cluster. 

The divide strategy based on the reflection $\mathbf{R}^\sigma$, designing an effective Hamiltonian $\bar{H}^c_{\rm eff}$ herein for the Hubbard model can be directly generalized to ab-initio Hamiltonians. The key distinction lies in ab-initio Hamiltonians in the presence of non-local electron-electron interactions, denoted as $U_{ijkl}$, which results in non-local correlation within the cluster Hamiltonian $\bar{H}^c_{\rm eff}$, and an additional contribution in $\bar{H}^{\rm env}_{\rm MF}$.

 Interestingly, as shown in \Fig{fig:schema_method}, concerning the {\it Divide} step, DMET entails performing the Schmidt decomposition of the wave function $\ket{\Psi}$. The exact projector $\mathbf{P}$ factoring $\ket{\Psi}$ is typically approximated by carrying out a singular value decomposition (SVD) of the elements of the 1-RDM connecting the fragment and the environment, which is exact when the wave function is mono-determinantal. The determined projector $\mathbf{P}$ allows for projecting the orbital subspace complementary to the fragment into an optimal subspace of the same dimension as the fragment (bath). This orbital subspace is identical to the one for the Householder transformation using the reflection $\mathbf{R}^\sigma$ ~\cite{sekaran_householder_2021,sekaran_unified_2023}.
Despite that $\mathbf{R}^\sigma$ keeps the fragment invariant, a mismatch can appear due to the approximation Eq.~(\ref{eq:DAC_eq}) between quantities computed either directly from $\gamma$ or using $|\bar{\Psi}^c\rangle$. 
Indeed, a mismatch appears already at the density matrix level, {\it i.e.} $ \tilde{\gamma}^\sigma|_c \neq \bar{\gamma}^\sigma$,  where $\bar{\gamma}^\sigma_{ij} = \langle \bar{\Psi}^c | \tilde{c}_{i\sigma}^{\dagger}\tilde{c}_{j\sigma} |\bar{\Psi}^c\rangle$ for $i$ and $j$ in the cluster. Equivalently the fragment kinetic energy can be computed as
\begin{eqnarray} \label{eq:Ekin_Frag}
      T_F = \sum_{i\in F}\sum_{j\sigma} t_{ij}\gamma_{ij\sigma} = \sum_{i\in F}\sum_{j\sigma}\tilde{t}^\sigma_{ij} \tilde{\gamma}^\sigma_{ij}
    \end{eqnarray}
    with $\tilde{t}^\sigma_{ij} =  \sum_{kl}\mathbf{R}^\sigma_{ik}t_{kl}\mathbf{R}^\sigma_{lj}$ and  differs from the fragment kinetic energy computed using $\bar{\gamma}^\sigma_{ij}$, that reads
    \begin{equation}
      \label{eq:Ekin_Frag_Clu}
      T_F^c =  \sum_{i\in F}\sum_{j\sigma}\tilde{t}^\sigma_{ij} \bar{\gamma}^\sigma_{ij}.
    \end{equation} 
 In order to overcome such mismatches, we propose the following iterative process where density matrices computed in clusters at iteration $s$, $\bar{\gamma}^{\sigma(s)}$, are used to reconstruct a new 1-RDM  of the full system for the $s+1$ iteration, $\gamma^{\sigma(s+1)}$.
 More precisely, following Eqs.~(\ref{eq:Ekin_Frag}) and~(\ref{eq:Ekin_Frag_Clu}), the reconstruction is performed using $\mathbf{R}^{\sigma(s)}$, the reflection determined at step $s$ such that for each fragment and each impurity $i$ in the fragment
\begin{flalign} 
 \gamma^{\sigma(s+1)}_{ij\sigma} &= \sum_{kl}\mathbf{R}^{\sigma(s)}_{ik}\bar{\gamma}^{\sigma(s)}_{kl}\mathbf{R}^{\sigma(s)}_{lj}  \quad \text{with }\mathbf{R}^{\sigma(s)}_{ik}= \delta_{ik}\\
&= \sum_{l\in \rm cluster}\bar{\gamma}^{\sigma(s)}_{il}\mathbf{R}^{\sigma(s)}_{lj} 
 \label{eq:trans_inv}
\end{flalign}
leading to  $T_F^{(s+1)} = {T_F^c}^{(s)}$.
Given that the fragment interaction energy $W_F$ is not an explicit functional of $\gamma$,   $W_F$  is assumed to be equal to the fragment interaction energy obtained in the cluster $W_F^c =\langle \bar{\Psi}^c |\bar{W}_F|\bar{\Psi}^c\rangle$.  It follows that  at convergence, the self-consistent reconstruction of $\gamma^\sigma$ leads to fulfill a local virial relation between the full system and each fragment, {\it i.e.}
     \begin{equation}
    \label{eq:Virial}
    \frac{T_F}{W_F} = \frac{T_F^c}{W_F^c},
  \end{equation}
providing a physically motivated justification for the 1-RDM {\it conquer} scheme. 
Drawing on the principles of the quantum virial theorem ~\cite{lowdin1959scaling}, the relation presented in \Eq{eq:Virial} presumes a universality in behavior among systems that exhibit analogous interactions.
For periodic systems, the translational invariance is not necessarily recovered when the fragment contains more than one impurity. To bypass this issue, we propose to modify \Eq{eq:trans_inv} as 
\begin{eqnarray} \label{eq:trans_inv2}
 \gamma^{\sigma(s+1)}_{ij} = \frac{1}{N_i}\sum_{m\in \rm frag}S_{im}\sum_{l\in \rm cluster}\bar{\gamma}^{\sigma(s)}_{ml}\mathbf{R}^\sigma_{lj},
\end{eqnarray}
where the operator $S_{im}$ shifts the impurity $m$ to the impurity $i$.
Inspired by Cluster Perturbation Theory proposed by Senechal \etal ~\cite{senechal_spectral_2000}, one can also (not done in this work) transform the lattice Hamiltonian such that the restriction to a finite cluster retains a periodic boundary within
the cluster, thus restoring the intracluster translational
symmetry.

We summarize the self-consistent algorithm that aims iteratively at defining a finite  1-RDM functional cluster Hamiltonian (\textit{Divide}) and to reconstruct full 1-RDM from the computation of local properties of fragments (\textit{Conquer}).
The algorithm starts at the iteration step $s=0$ with a trial 1-RDM $\gamma^{(s)}$ (such as obtained within the Hartree-Fock approximation for example) and a given partition into fragments.\\
 For each spin $\sigma$ and fragment $F$ do:
 
{\it (i)} Determine the unitary transformation $\mathbf{R}^{\sigma (s)}[\gamma^{\sigma (s)}]$ and compute the Hamiltonian $\tilde{H}$ in the $\mathbf{R}^{(s)}$ representation using the transformations~(\ref{eq:transfo1}).

{\it (ii)} Define the density matrix functional effective cluster Hamiltonian $\bar{H}^{c(s)}_{\rm eff}$ following Eq.~(\ref{eq:cluster_Ham}). Diagonalize $\bar{H}^{c(s)}_{\rm eff}$ to obtain $|\bar{\Psi}^{c(s)}\rangle$ and local impurity fragment properties such as the cluster 1-RDM $\bar{\gamma}^{\sigma (s)} = \langle \bar{\Psi}^{c(s)} | \hat{\gamma}^\sigma| \bar{\Psi}^{c(s)} \rangle$.

{\it (iii)} Use \Eq{eq:trans_inv} or \Eq{eq:trans_inv2} to construct the 1-RDM of the full system.

{\it (iv)} If the virial relation in \Eq{eq:Virial} is not reached, update the iteration step with the new 1-RDM and go back to step {\it (i)} until convergence is achieved for each spin and fragment. 
Note that to avoid drastic changes of the 1-RDM and convergence issues, a damping parameter $d$ is added, meaning that a fraction $d$\ of the previous density matrix obtained is kept in the new density-matrix. More sophisticated and efficient damping algorithms might be used, e.g. in the spirit of the Direct Inversion in the Iterative Subspace (DIIS) technique ~\cite{pulay_convergence_1980}. In the results presented here we used $d = 60\%$.
Finally, the process halts if an updated 1RDM becomes non N-representable, assuming that the minimal virial condition has been achieved within the area of representability.
As an example, we analytically characterize the accessible 1-RDM domain by the reconstruction for the single impurity case in Appendix~\ref{app:single_imp}.
  \subsection{{\it Conquer} versus {\it Matching}}
  \label{sec:cDMET}
  
 \begin{figure}
    \centering
    \resizebox{\columnwidth}{!}{
      \includegraphics[scale=0.80]{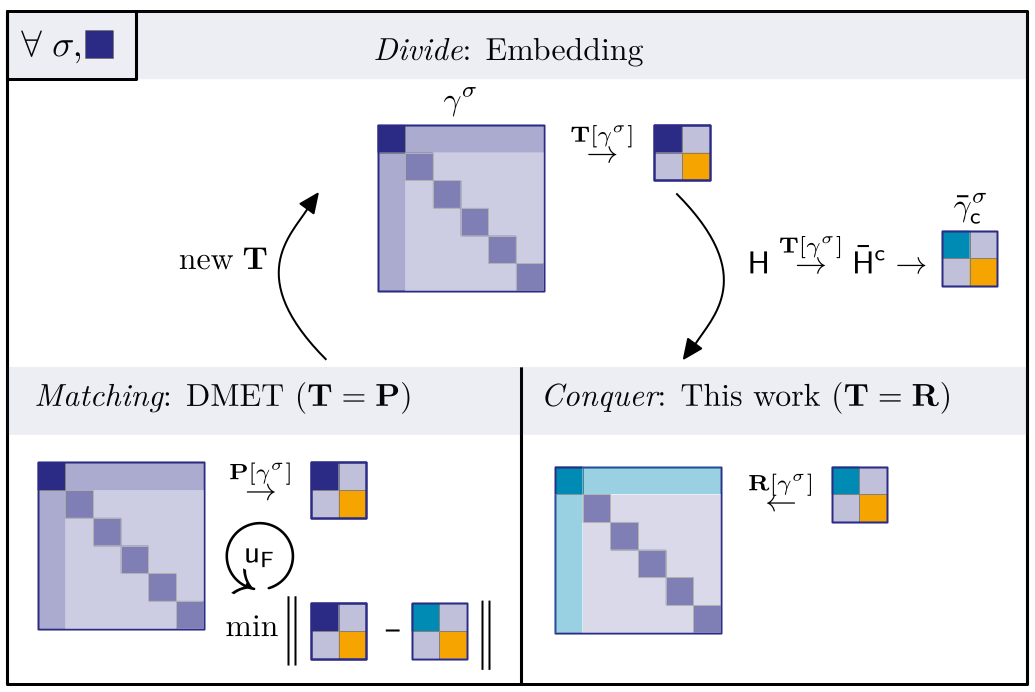}
    }
    \caption{Representation of the Divide-and-Conquer protocol proposed in this work compared to the standard DMET self-consistent scheme.}
    \label{fig:schema_method}
  \end{figure}
  In this section, we discuss the advantages of performing the {\it Conquer} approach used in DaC rather than the {\it Matching} procedure of DMET. 
As schematized in \Fig{fig:schema_method}, the conquest in the present work aims at reconstructing the global 1-RDM by performing the inverse transformation of all computed cluster density matrices $\bar{\gamma}$, by using the well-defined unitary reflection $\mathbf{R}^\sigma$. 
In DMET, the projector $\mathbf{P}$ is used to define the effective subspace. Because it is not unitary, it leads to a reduction of the full space into an smaller subspace. Moreover it seeks an effective potential $u_F$ through a numerical optimization method, such that the idempotent projected 1-RDM achieves optimal overlap with the cluster density matrix. Additionally, different matching protocols have been proposed ~\cite{bulik_density_2014}. The matching protocol used can drastically influence the physical properties obtained through the self-consistent scheme, such as the band gap at half-filling for the 1D-Hubbard model, or the unphysical chemical potential as discussed in Ref. \cite{knizia_density_2012}.
The current DaC algorithm allows for the direct reconstruction of the density matrix via \Eq{eq:trans_inv}, thereby enabling the immediate updating of 1-RDM fragment quantities in the full system without adding any effective potential.
 The DMET approach, in practice, does not enable direct reconstruction of the density matrix. When utilizing a ``democratic'' representation~\cite{wouters_practical_2016}, the obtained 1-RDM suffers from significant representability issues~\cite{wu_projected_2019, nusspickel_effective_2022}. More sophisticated approaches exist and are referred as non-democratic~\cite{nusspickel_effective_2022}.
  \section{Results \& Discussion} 
In this section, the DaC algorithm is applied to large paramagnetic Hubbard rings with $N=414$ sites and compared with the Bethe-Ansatz (BA)~\cite{lieb_absence_1968} and standard DMET~\cite{knizia_density_2012}. DMET calculations have been achieved by considering the explicit two-body interactions in the bath, the mean-field correction proposed in \Eq{eq:mf_ce}, and matching the fragment density matrix.
\begin{figure}
  \centering
  \includegraphics[scale=0.25]{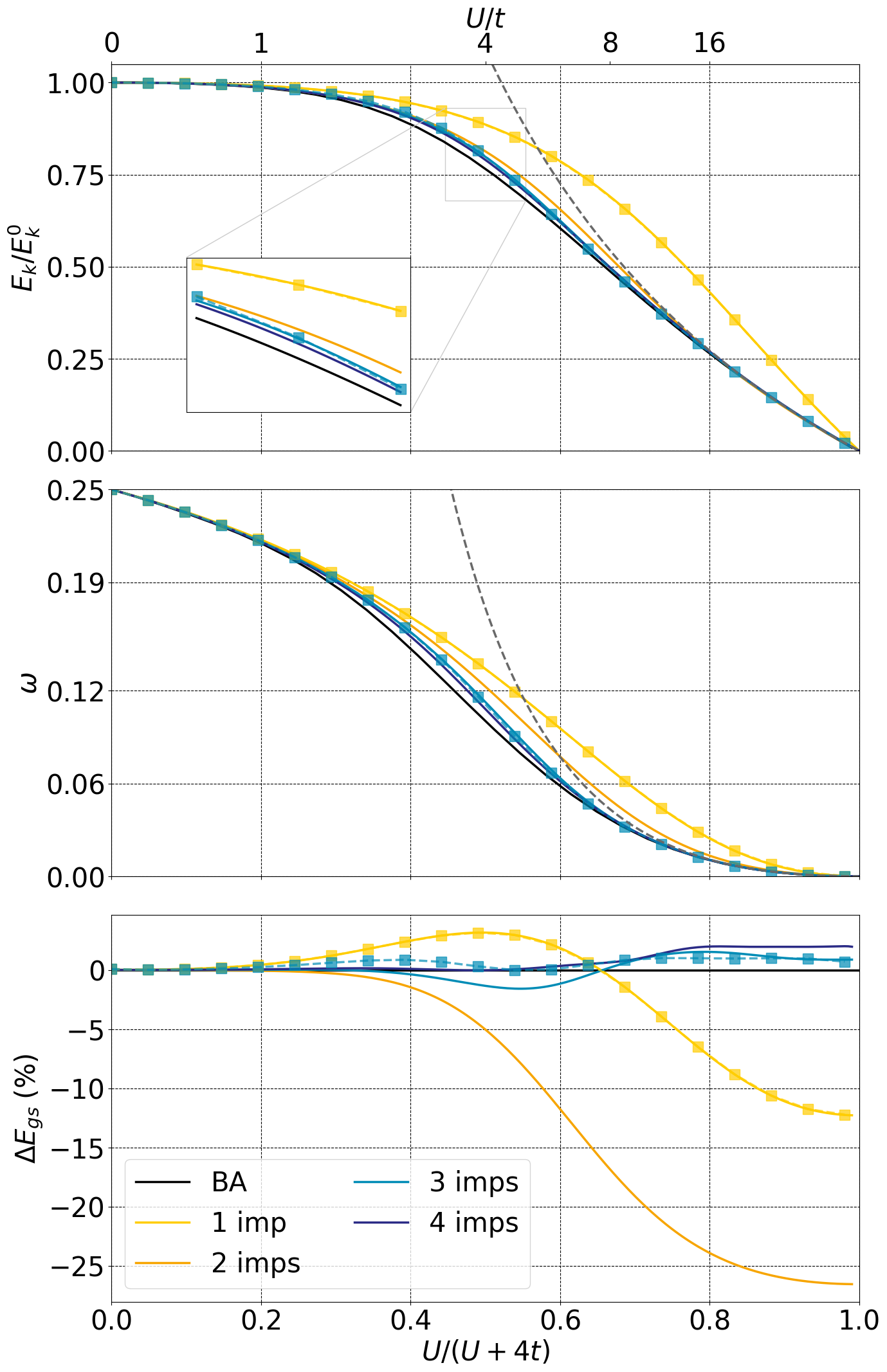}\\ 
  \caption{Kinetic energy renormalized with the non-interacting kinetic energy $E^0_k=-4N_s/\pi$ (top panel), average double occupation per site $\omega$ (middle panel) and relative error on the ground state energy $\Delta E_{gs} = 100\times(E_{gs} - E_{\rm BA})/E_{\rm BA}$ (bottom panel) for 1D half-filled Hubbard ring with respect to the relative strength of the Coulomb repulsion $U/t$. Results are provided for different numbers of impurities (full lines). Comparison is made with the exact Bethe Ansatz (black solid line).  Squares and dashed lines correspond to standard interacting-bath DMET. Dark gray dashed lines correspond to the asymptotic behavior in the strongly correlated limit.}
  \label{fig:energies}
\end{figure}
Let us first focus on the less trivial half band filling case $N_s = N_e$, $N_e$ being the total number of electrons in the system. In \Fig{fig:energies} we show results for energies as a function of the relative strength of the Coulomb repulsion $U/t$. More precisely we display the rescaled kinetic energy $E_k/E_k^0$ where $E_k =-t \sum_{\langle i,j \rangle \sigma}\gamma^\sigma_{ij}<0$ and $E_k^0$ is the non-interacting kinetic energy, the average double-occupation number per site $\omega = \langle \hat{n}_{i\uparrow}\hat{n}_{i\downarrow} \rangle$ and the relative ground-state energy  error $\Delta E_{gs}$ with respect to the energy obtained with BA. 
Concerning asymptotic behaviors, both the kinetic energy and the
double occupation are well reproduced in the non-interacting limit.
In contrast, as for DMET and in the strongly-interacting limit, asymptotic behaviors are recovered only for impurity numbers in the fragment larger than two.

As depicted in the top (middle) panel of \Fig{fig:energies}, the kinetic energy ratio and the double-occupation are both significantly overestimated in comparison to the exact reference for the single impurity case (yellow line).
Consequently, the ground state energy corresponding to the sum of previous contributions, namely $E_{gs} = E_k + U \omega$ benefits from significant error compensation, resulting in a relative error below $4\%$ for $U/t<8$.
In addition, as detailed Appendix~\ref{app:single_imp}, self-consistency is achieved  at the first iteration without damping and leads exactly to the same energy as DMET calculation (yellow squares).
Increasing the number of impurities up to four drastically reduces the error compensation. In particular, in the strongly correlated regime $U \gg t$, the kinetic energy (double occupation) follows the exact asymptotic behavior equal to $-8$ln($2$)$t^2/U$ ($4$ln($2$)$t^2/U^2$), respectively~\cite{lopez-sandoval_density-matrix_2002}.
The systematic improvement of the kinetic energy and the double occupation as the number of impurities increases demonstrates the relevance of the proposed embedding strategy. This is consistent with the convergence of other embedding approches such as DMET with respect to the number of impurities~\cite{lee_accuracy_of_2023}.
Ultimately, ground-state energy errors smaller than one percent are obtained already for three impurities. Note that at large values of $U/t$, precision error and convergence issues occur since the density matrix is close to be diagonal.
  \begin{figure}
\centering
    \resizebox{\columnwidth}{!}{
    \includegraphics[scale=0.12]{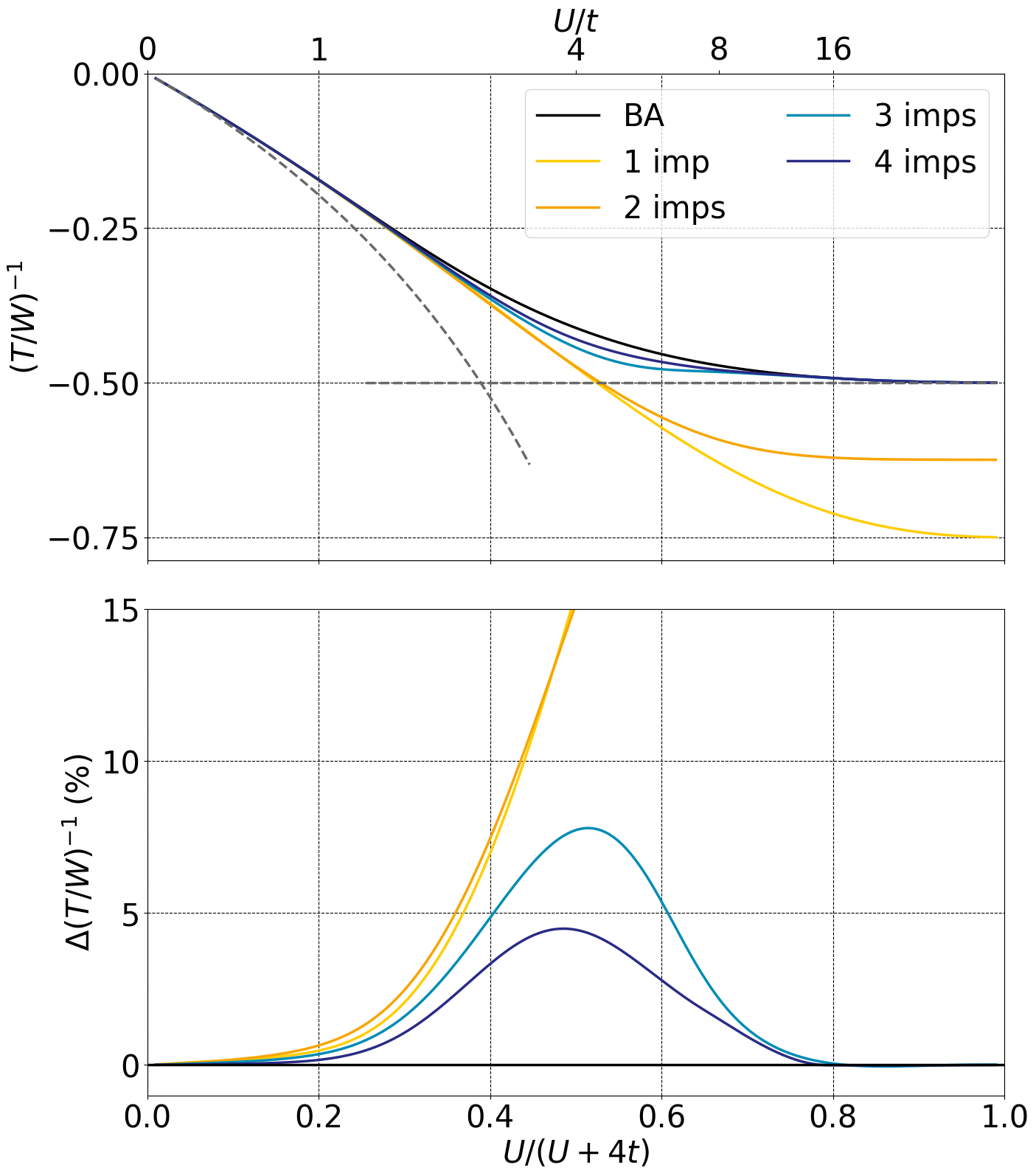}
    }
    \caption{
Ratio between interaction and kinetic energies $W/T$ (top panel) and associated relative error $\Delta(W/T)$ (bottom panel) with respect to the correlation strength $U/t$ and up to 4 impurities in the fragment.  Results (full colored lines) are compared with the Bethe Ansatz solutions (black line) and asymptotic behaviors in the non- and strong-correlated limits are highlighted in dark gray dashed lines.}
    \label{fig:Virial}
  \end{figure}
The ratio between interaction and kinetic energies $W/T$ and the associated relative error $\Delta (W/T)$ are shown in \Fig{fig:Virial} with respect to the interaction strenght $U/t$ and for different fragment sizes. 
Starting from $U/t=0$, $W/T$ decreases asymptotically as $f(U/t)=-U\pi/(16t)$ (mean-field limit) and monotonously converges in the strongly interacting limit to $W/T=-1/2$. 
The strongly correlated limit of $W/T$ depends on the topology of the system. More precisely, all systems where the energy of the ground state exhibits an asymptotic behavior proportional to $t^2/U$, such as the 1D Hubbard model, have a limit of $W/T$ equal to $-1/2$.
Indeed, at this limit, with the Hellmann-Feynman theorem, we obtain\\
\begin{flalign}
T &= t\times \partial E_{gs} / \partial t = 2E_{gs},\\
W &= U \times \partial E_{gs} / \partial U = -E_{gs}.
\end{flalign}
In this work $T/W$ serves as a metric of the self-consistent condition since $T/W$ is enforced to be equivalent in both the extended system and the embedding cluster. 
Consequently, the error $\Delta (W/T)$ appears as a measure of the embedding efficiency, {\it i.e.} the ability of the embedded impurity to mimic the extended system. As shown in the bottom panel of \Fig{fig:Virial},  $\Delta (W/T)$  systematically decreases as the number of impurity increases and, importantly, the strongly correlated asymptotic behavior is correctly reproduced from three impurities.
This highlights that clusters obtained by considering  one and two impurities are too small to reproduce correctly charge and spin fluctuations of the extended systems at large Coulomb interaction strength. 
 Ultimately, errors obtained  for three and four impurities are maximal in intermediate correlated regime and become lower than  $5\;\%$ for four impurities.
 \begin{figure}
 \centering
     \resizebox{\columnwidth}{!}{
     \includegraphics[scale=0.12]{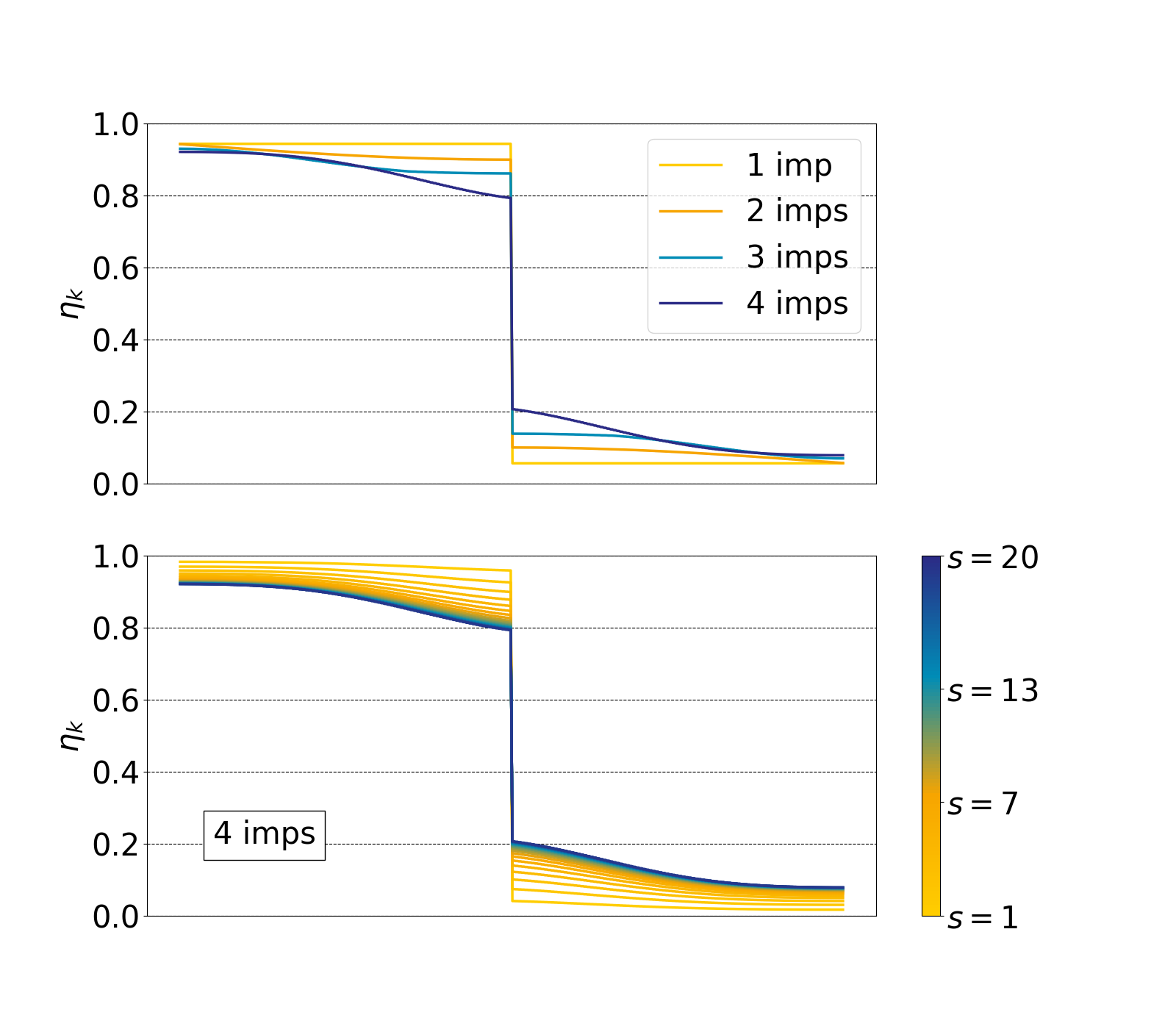}
     }
     \caption{ Natural orbital occupancy profile in descending order at correlation strength $U/t=4$ for different impurity numbers in the fragment (top panel) and as a function of the iteration number $s$ as displayed by the colorbar (bottom panel). The x-axis refers to the natural orbital index.}
     \label{fig:eta_k_full}
 \end{figure} 
As mentioned in \Sec{sec:cDMET}, a major difference between our work and DMET consists in the ability to obtain full size non-idempotent density matrices.
  This peculiarity is highlighted in \Fig{fig:eta_k_full} where the occupation number $\eta_k$ of natural  orbitals are shown in descending order for $U/t = 4$ and impurity numbers from one to four.
 Occupation numbers  $\eta_k$ and natural orbitals are calculated by diagonalizing the converged 1-RDM obtained by using \Eq{eq:trans_inv2}. Natural orbitals obtained are Bloch states (not shown), {\it i.e.} translational invariance of the system is restored at the end of the self-consistent scheme.
  Importantly, all converged 1-RDM are shown to be N-representable, {\it i.e.} with occupation number in between zero and one. For the single impurity case,  occupation numbers behave as a step function, as discussed in Appendix ~\ref{app:single_imp}.
  By increasing the number of impurities, profiles become much more complex but still smooth.
  Note that  whatever the impurity number, the particle-hole symmetry expected at half-filling is fulfilled.
  
  The self-consistent process is illustrated in \Fig{fig:eta_k_full} that displays natural orbital occupation numbers in descending order for $U/t = 4$ and four impurities at every iteration step up to twenty iterations.
  This highlights a smooth convergence and that at each iteration $s$ the density matrix $\gamma^{(s)}$ also belongs to the N-representability domain, and preserves particle-hole symmetry.
  
\begin{figure}
\centering
    \resizebox{\columnwidth}{!}{
    \includegraphics[scale=0.12]{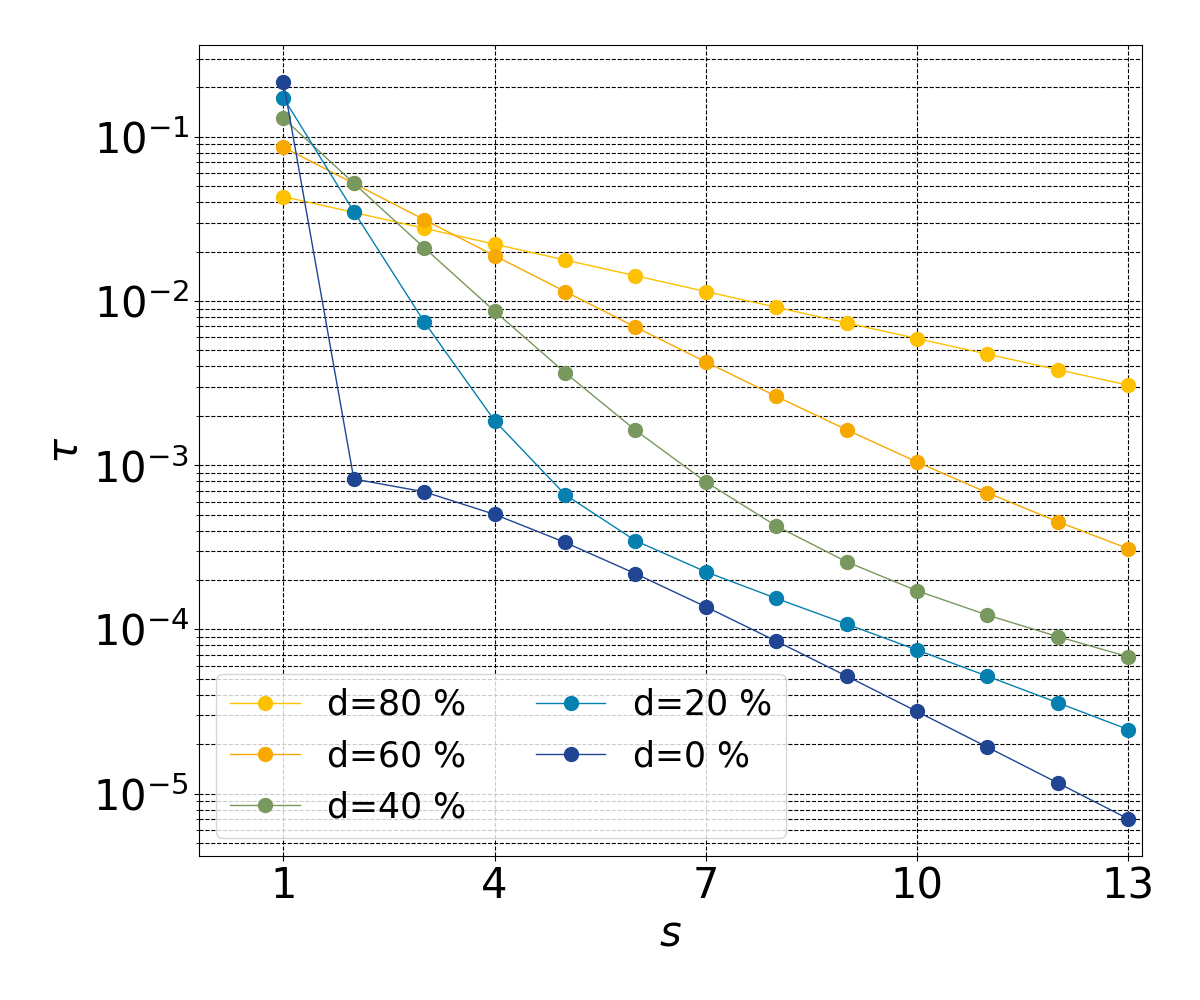}
    }
    \caption{Evaluation of the virial convergence parameter defined in \Eq{eq:Virial} $\tau(s) = \left(\frac{T_F}{W_F} - \frac{T_F^c}{W_F^c}\right)^{(s)}$ as a function of the number of steps $s$ of the self-consistent scheme for $U/t=4$. Results are provided for different number of dampings. Lines are guidelines for better visualization}
    \label{fig:damping}
  \end{figure}
  
We have observed that the iterative process allows electronic correlation to be taken into account in the extended system. For that, it is necessary to identify a fixed point in the divide and conquer application detailed in \Sec{sec:sces}, implying that the newly formed density matrix is identical to the initial one. We have also proposed a self-consistent approach to determine this fixed point. However, there are no relationships that exist to demonstrate that this fixed point is existent, unique, and consistently reached.
Thus, we define the virial convergence parameter $\tau(s) = \left(\frac{T_F}{W_F} - \frac{T_F^c}{W_F^c}\right)^{(s)}$ as  the difference between fragments $T/W$ ratio in the extended system and in the cluster calculated at steps $s$.  
The introduced convergence parameter enables an evaluation of the embedding scheme at each iteration $s$, providing a quantitative understanding of the proposed self-consistent process.
  In \Fig{fig:damping} we present $\tau(s)$ as a function of the number of iteration $s$ of the self-consistent scheme for the interaction strength $U/t=4$, two impurities in the fragment and different damping parameters $d$.
For large damping parameter ($d>50\%$), the convergence is smooth and $\tau(s) \propto e^{as+b}$ follows an exponential law (linear in logarithmic scale) where the convergence velocity $a$ increases with respect to $d$.
For low or zero damping 
parameters, instabilities of the the self-consistent scheme can be observed during the first iterations. 
Instabilities are associated with large value of $\tau$, of the same order than $T/W$ and an  abrupt and non-monotonous decrease of $\tau$ as a function of $s$. 
Consequently, a compromise  between fast convergence and stability of the process has to be reached.
\begin{figure}
\centering
    \resizebox{\columnwidth}{!}{
    \includegraphics[scale=0.12]{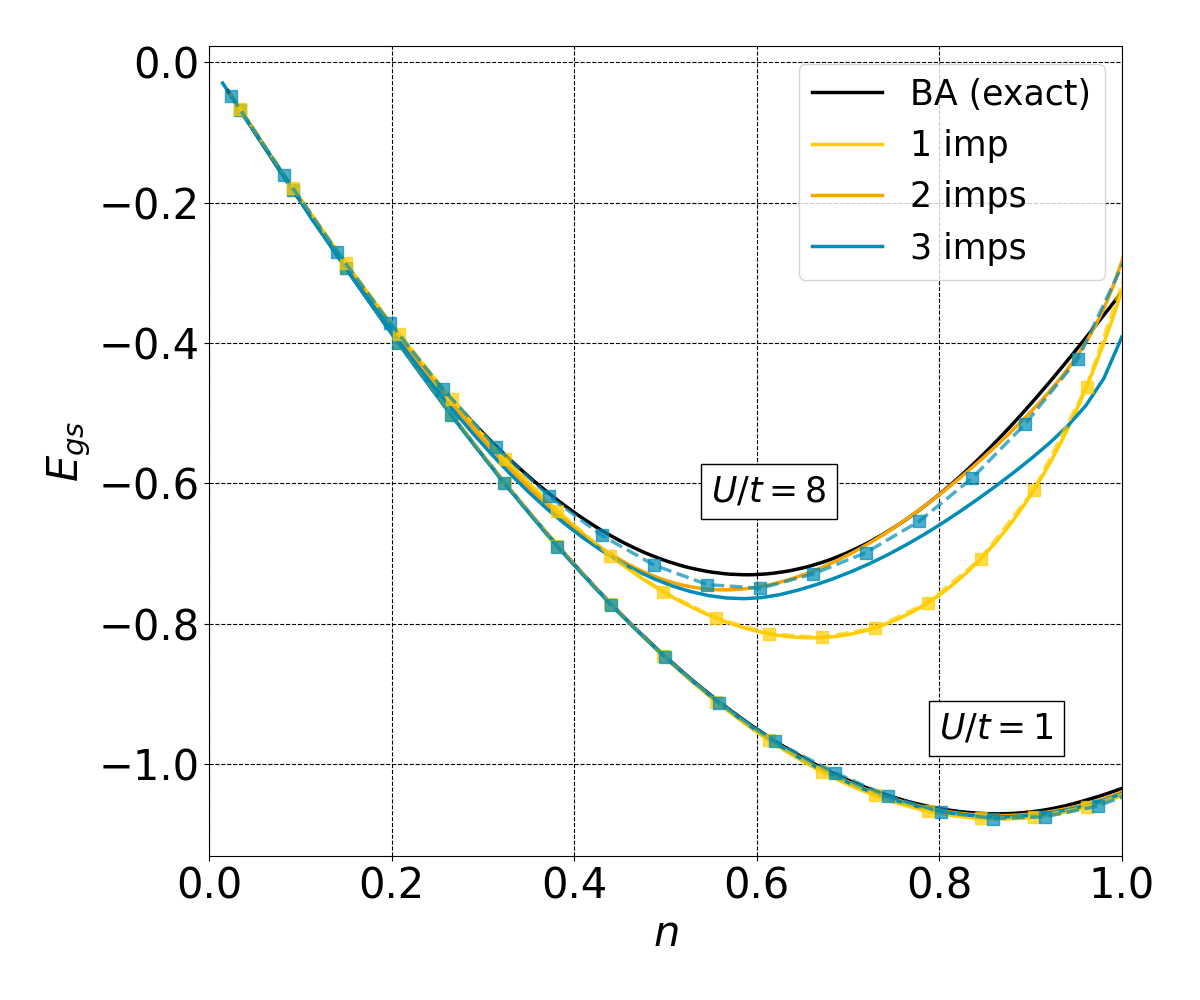}
    }
    \caption{Ground state energy calculated with respect to the filling n for $U/t=1$ (low correlated regime) and $U/t=8$ (strongly correlated regime). Results are provided for different numbers of impurities (full lines). Comparison is made with the exact Bethe Ansatz (black solid line).  Squares and dashed lines correspond to interacting-bath DMET.}
    \label{fig:e0_fill}
  \end{figure}
  
\begin{figure}
\centering
    \resizebox{\columnwidth}{!}{
    \includegraphics[scale=0.12]{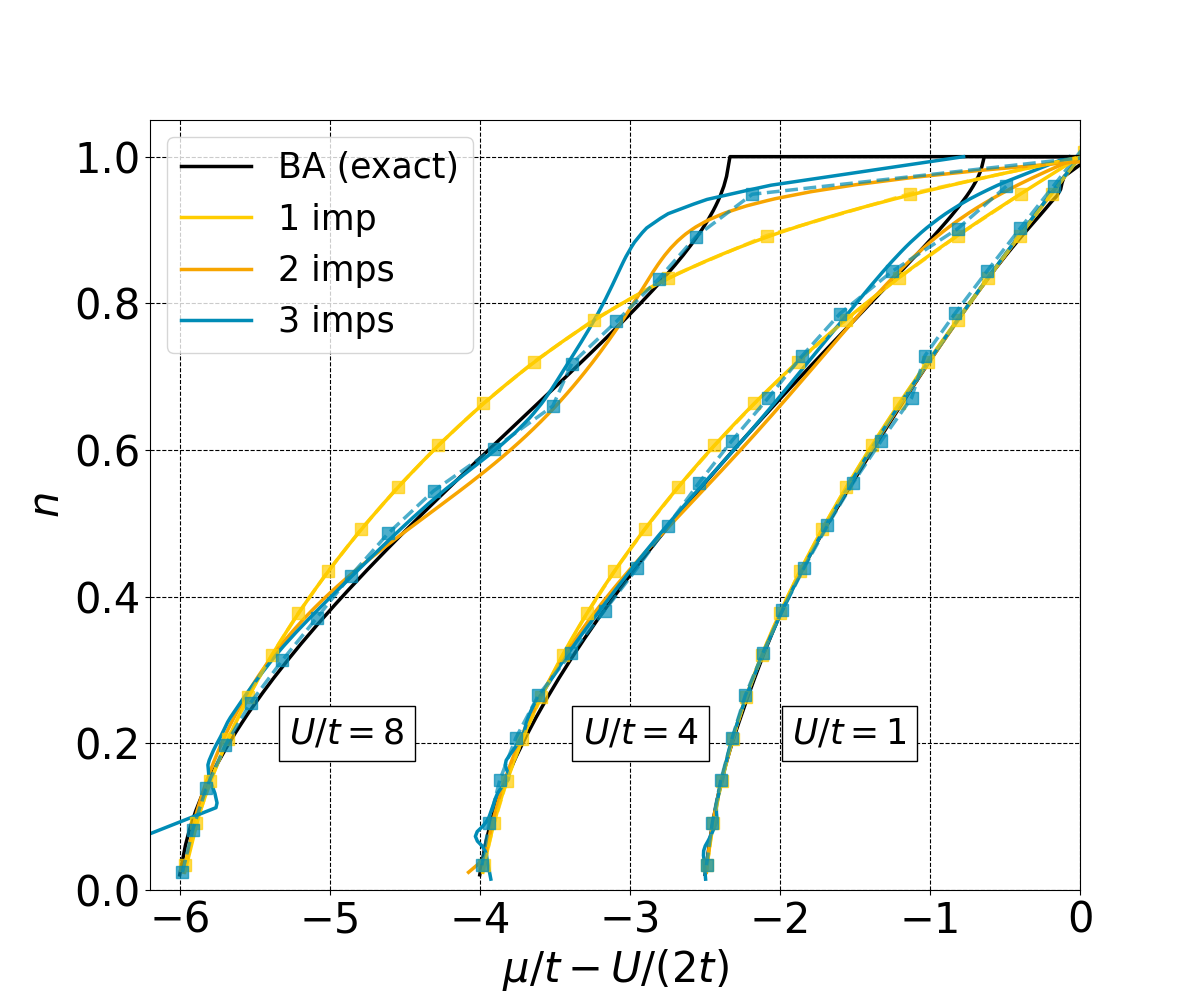}
    }
    \caption{Lattice filling $n$ with respect to the chemical potential $\mu(n) = \partial E_{gs} / \partial n$ calculated for $U/t=1,4,8$. Results are provided for different numbers of impurities (full lines). Comparison is made with the exact Bethe Ansatz (black solid line).  Squares and dashed lines correspond to interacting-bath DMET.}
    \label{fig:mu}
  \end{figure}
Let us focus on filling $n=N_e/N_s$ away from one and in particular for $0 \leq n \leq 1$ due to electron-hole symmetry. 
We present in \Fig{fig:e0_fill} the ground-state energy $E_{gs}$ as a function of $n$ considering one to three impurities in the fragment and in the weakly (strongly) interacting regime $U/t =1$ ($U/t =8$), respectively.
For $U/t=1$, the ground-state energy is correctly reproduced for any filling. For $U/t=8$ the single impurity case underestimates the ground-state energy up to within $16\%$ error for $n=0.7$.
While increasing the number of impurities improves the single impurity results, interestingly, the energy for three impurities appears worse than the two impurities case due to error compensation. 
Indeed, the following the results for the half-filled case in \Fig{fig:energies}, both the kinetic and the interaction energies are individually improved going from two to three impurities in the fragment (not shown).
To investigate the insulating versus metal character , we present in \Fig{fig:mu} the chemical potential $\mu = \partial E_{gs}/\partial n$ as a function of $n$ for different values of the Coulomb interaction relative strength $U/t = 1,4,8$ and different sizes of impurity fragments.
Exact results are well reproduced both within our approach or with DMET, whatever the size of the impurity fragment in the weakly interacting regime ($U/t = 1$), or for low filling, even though some numerical instabilities are observed ($n < 0.2$). Importantly, no insulating regime is observed at half-band filling within our approach or DMET, even though improvement of the shape is observed for $n>0.8$.
\begin{figure}
\centering
    \resizebox{\columnwidth}{!}{
    \includegraphics[scale=0.12]{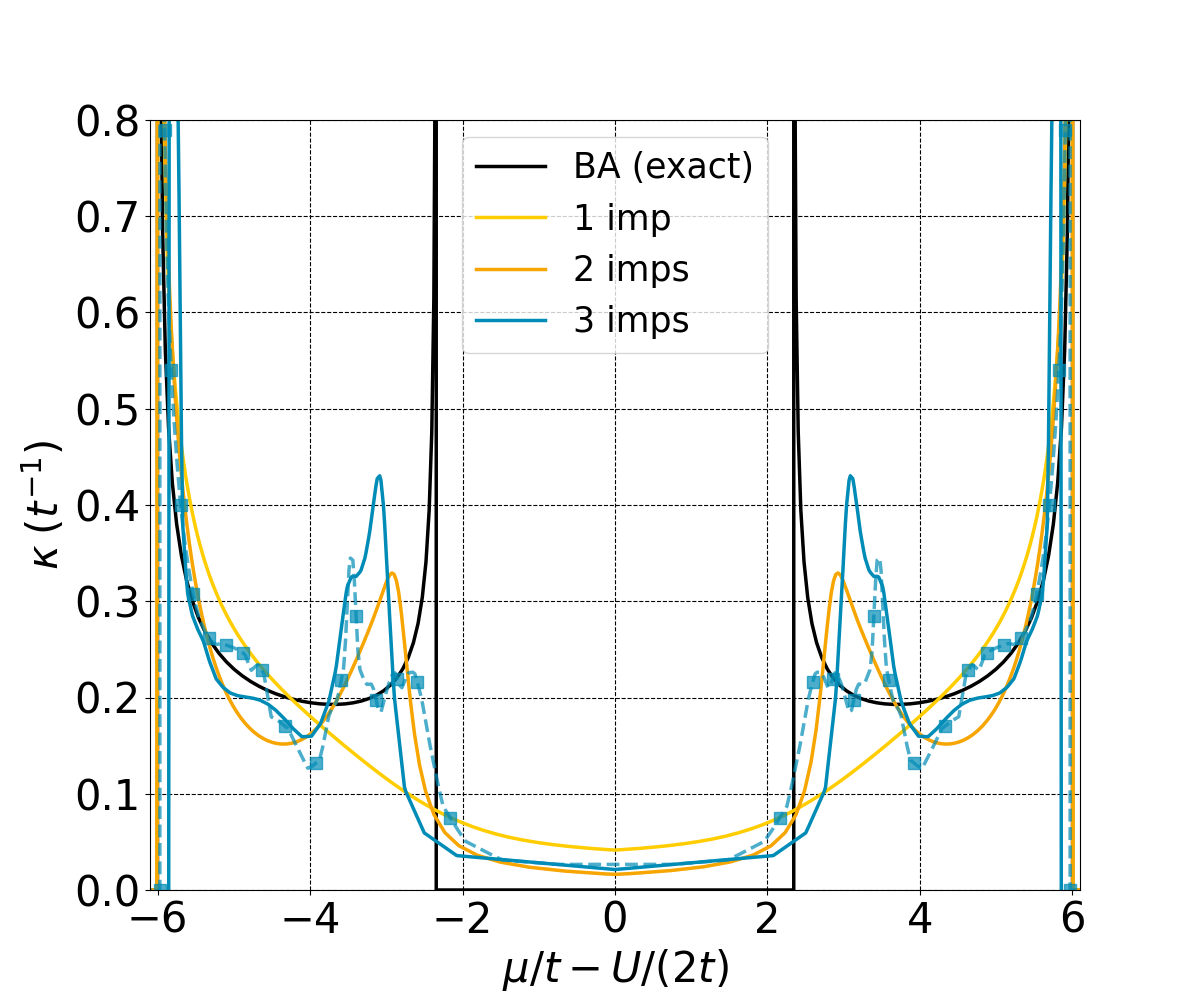}
    }
    \caption{Charge compressibility $\kappa = \partial n/ \partial \mu$  with respect to the chemical potential $\mu$ calculated for $U/t=8$. Results are provided for different numbers of impurities (full lines). Comparison is made with the exact Bethe Ansatz (black solid line).  Squares and dashed blue line correspond to three impurities interacting-bath DMET.}
    \label{fig:chi}
  \end{figure}
The Mott transition can be further investigated by looking at the charge compressibility presented in \Fig{fig:chi} as a function of the chemical potential $\mu$.
    The charge compressibility $\kappa = \partial n/ \partial \mu$ 
    refers to the response of the system's electron density to an external perturbation, such as an applied electric field.
    Specifically, it measures the degree to which the electron density fluctuates in response to a small change in the chemical potential, which is a measure of the energy required to add or remove an electron from the system.
    Without interaction $\kappa$ is equivalent to the one-body density of states of the corresponding non-interacting system.
    For the Bethe Ansatz and at finite $U/t$ the Mott gap opens at half band, splitting the band between the so-called lower and upper Hubbard bands that correspond to hole and electron doping, respectively.
    Moreover, at the band edges close to half-band filling, for $\mu(n=1^-)$ and  $\mu(n=1^+)$, $\kappa$ sharply diverges.
    As already discussed, both the presented DaC algorithm and DMET fail to reproduce the band splitting as well as the divergences close to half-band filling.
    However, by increasing the number of impurities in the fragment, band splitting, even if non fully achieved, becomes clearly visible and the divergence close to half-filling becomes sharper for three impurities. Note that as the number of impurities increases from two to three, the main peak position corresponding to $\mu(n=1^-)$ ($\mu(n=1^+)$) shifts to lower (higher) energy, respectively. We suspect that this phenomenon may arise either due to error compensation in the ground state energy or could be attributed to oscillations in the amplitudes of the charge gap within finite-sized clusters.   
    While the width of the pseudo sub-bands are better reproduced by DMET with three impurities, DMET also shows a wavy behavior in the sub-bands that arises from the many inflection points of the filling versus chemical potential, see \Fig{fig:mu}.
    Note that the matching process used in DMET can also lead to drastic changes in the compressibility plot. Indeed, matching 1-RDMs over the full cluster would lead to unphysical infinitely negative charge compressibility at the chemical potential for which a bending is observed in the density versus chemical potential plot~\cite{knizia_density_2012}.
    All together, charge compressibility results are quite encouraging for our DaC algorithm as the shape of the compressibility close to half-band filling improves by increasing the number of impurities demonstrating the efficiency of the embedding to account for low-energy charge and spin fluctuations.
  
\section{Conclusions}
In this paper, we propose a self-consistent quantum embedding scheme based on the reduced density matrix formalism and tested against the 1D Hubbard model. 
This original approach allows to obtain all the reduced quantities of the system extracted from a smaller cluster solved with high accuracy. 
This is made feasible through the use of a unitary transformation of the Hamiltonian via a involution $\mathbf{R}^\sigma[\gamma^\sigma]$ determined self-consistently .
The accuracy of computed properties depends directly on the size of the cluster, where increasing the size of the cluster leads systematically to an improvement of the results. 
Self-consistency is achieved by using virial-like relationship, which provides quantitative insight of the correspondence between the fragment in the extended system and its embedded counterpart. 
In contrast to DMET, our approach needs no external, nonlocal and ad-hoc potential to realize the self-consistent scheme and leads to an N-representable reconstruction of the correlated 1-RDM.
The encouraging results presented in this contribution, motivate further applications on systems beyond the 1D case.
In a future work, other systems, such as the 2D Hubbard model will be explored. In addition, other single-particle reduced quantities will be studied, such as dynamical quantities with Green's function $G(iw)$ or self-energy $\Sigma(iw)$, and will also be examined and contrasted with state-of-the-art methods such as DMFT ~\cite{georges_dynamical_1996}, or the energy-weighted DMET ~\cite{fertitta_energy-weighted_2019}.
Two-particle response functions will also be studied, such as the non-local two-particle reduced density matrix $\Gamma$, where there is no systematic way to extrapolate from the cluster to the system without breaking fermionic symmetries ~\cite{nusspickel_effective_2022}, or the spin response function $\langle S_iS_j \rangle$. 
Dynamical two-particle response functions can also be extracted, such as the charge/spin polarization used in the Bethe-Salpeter equations for example.
\begin{acknowledgments}
The authors would like to thank the ANR (Grant No. ANR-19-CE29-0002 DESCARTES
project) for funding.
E.F. thanks the ANR (CoLab project, grant no.: ANR- 19-CE07-0024-02) and the University of Strasbourg (IdEx 2021 call, Grant No. W21RPD04) for funding.
\end{acknowledgments}

\appendix

\section{Construction of the reflection}\label{app:reflection}
The unitary reflection functional of the 1-RDM $\mathbf{R}^\sigma[\gamma]^\sigma$ is designed to fulfill properties  defined in \Eq{eq:impkeep} and \Eq{eq:dis}.
For clarity, the spin index $\sigma$ is omitted for the reflection matrix $\mathbf{R}^\sigma$ and the spin 1-RDM $\gamma^\sigma$.
For that, we propose to follow the work of Rotella ~\cite{rotella_applied_nodate} in order to construct the reflection $\mathbf{R}$. 
Let first consider the matrix $\mathbf{V}\in\mathcal{M}_{Nr}(\mathbb{R})$, where $N$ correspond to the number of orbitals and $r \leq N/2$ the rank of the matrix (equivalently the number of impurity(ies) in the fragment), we define $\mathbf{R}$ by\\
\begin{eqnarray}  \label{eq:reflection}
\mathbf{R} = \mathbb{1} - 2\mathbf{V}(\mathbf{V}^T\mathbf{V})^{-1}\mathbf{V}^T.
\end{eqnarray} 
By construction, the matrix $\mathbf{R}$ is an involution, {\it i.e.} $\mathbf{R}^{-1}=\mathbf{R}$. As explained in \Sec{sec:emb}, this can be geometrically interpreted as a reflection.
The orthonormalization part $(\mathbf{V}^T\mathbf{V})^{-1}$ is here to ensure the unitarity of the matrix $\mathbf{R}$.
The matrix $\mathbf{V}$ is constructed as following\\
\begin{eqnarray}
\mathbf{V} = \left(
\begin{array}{c}
\mathbf{0}_{r} \\
\mathbf{A}_1 + X \\
\mathbf{A}_2
\end{array}
\right)
\end{eqnarray}
The matrix $\mathbf{0}_{r} \in\mathcal{M}_{rr}(\mathbb{R}) $ is the null matrix, that impose identity over impurity(ies) (see \Eq{eq:impkeep}). Matrices $\mathbf{A}_{1} \in\mathcal{M}_{rr}(\mathbb{R}) $ and $\mathbf{A}_{2} \in\mathcal{M}_{(N-2r)r}(\mathbb{R}) $ are parts of the matrix $\mathbf{A}$ corresponding to $r$ first columns of the 1-RDM $\gamma$\\
\begin{eqnarray} \label{eq:gamma}
\mathbf{A} = \left(
\begin{array}{c}
\mathbf{A}_{0} \\
\mathbf{A}_1 \\
\mathbf{A}_2
\end{array}
\right)
\end{eqnarray}
Finally, the matrix $\mathbf{X} \in\mathcal{M}_{rr}$ is designed to impose property defined in \Eq{eq:dis}. In that way, $\mathbf{X}$ is constructed such that\\
\begin{eqnarray}  \label{eq:prop2}
\mathbf{R}\gamma = \left(
\begin{array}{c}
\mathbf{A}_{0} \\
\tilde{\mathbf{A}}_1 \\
\mathbf{0}_{r}
\end{array}
\right)
\end{eqnarray}
Inserting \Eqss{eq:reflection}{eq:gamma} in \Eq{eq:prop2} yields:\\
\begin{eqnarray}\label{eq:Xprop}
\mathbf{X}^T\mathbf{X} = \mathbf{A}^T\mathbf{A}
\end{eqnarray}
where $\mathbf{X}$ is such that $\mathbf{A}^T_1\mathbf{X}$ is symmetric. 
Note that the solution of \Eq{eq:Xprop} is not necessarily unique.  
Rotella ~\cite{rotella_applied_nodate} proposed a systematic way to solve such constrained equation if $\mathbf{A}_1$ is non-singular.
 Let us introduce the matrix $\mathbf{Z} = \mathbf{X}\mathbf{A}^{-1}_1$ which must also be symmetric. We can write the previous identity as \\
\begin{eqnarray}
\mathbf{Z}^2 = \mathbb{1}_r + \mathbf{\Lambda}^T\mathbf{\Lambda}
\end{eqnarray}
where $\mathbf{\Lambda} = \mathbf{A}_2\mathbf{A}^{-1}_1$. Because $\mathbf{Z}^2$ is positive definite, we obtain the square root with\\
\begin{eqnarray}
\mathbf{Z} = \mathbf{P}^T\sqrt{\mathbf{D}}\mathbf{P}
\end{eqnarray}
With $\mathbf{P}$ is the matrix of orthogonal eigenvectors of $\mathbf{Z}^2$ and $\mathbf{D}$ the matrix of eigenvalues.
Finally, we obtain the solution\\
\begin{eqnarray}
\mathbf{X} = \mathbf{P}^T \sqrt{\mathbf{D}}\mathbf{P}\mathbf{A}_1
\end{eqnarray}
leading to the correct construction in \Eq{eq:prop2} with $\tilde{\mathbf{A}}_1 = -\mathbf{X}$. 
\section{Single impurity case} \label{app:single_imp}
\begin{figure}
\centering
    \resizebox{\columnwidth}{!}{
    \includegraphics[scale=0.12]{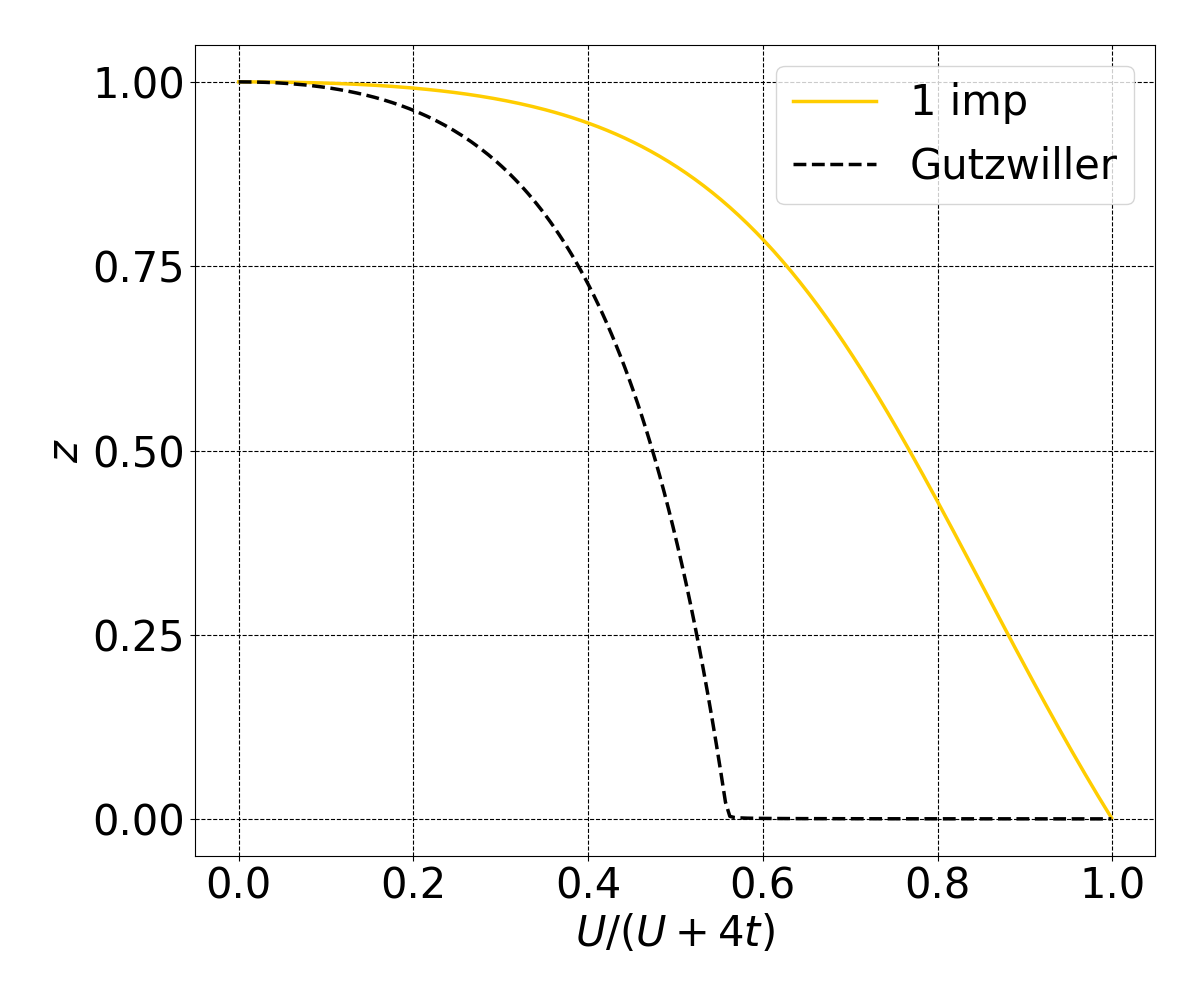}
    }
    \caption{Renormalization factor $z$ for the single impurity case (yellow line) compared to the Gutzwiller approximation (black dashed line) with respect to the correlation strength $U/t$.}
    \label{fig:single_imp}
  \end{figure}

The divide and conquer process proposed in \Sec{sec:sces} concludes in a single iteration and can be analytically determined in the single impurity case.
The ``divide'' (or embedding) step has been highly investigated in this case in the work of Sekaran \etal{}~\cite{sekaran_householder_2021}. 
In this part, we focus on particular properties of the ``conquer'', self-consistent scheme for the single impurity case. 
We assume that the self-consistent scheme starts with the non-interacting (idempotent) density matrix $\gamma^{\sigma(s=0)}$, where superscript $s$ refers to the self-consistent step.
In this simple case, elements of the unitary reflection become $\mathbf{R}^{\sigma(0)}[\gamma^{\sigma(0)}]_{ij} = \delta_{ij} - 2v_iv_j$ with \\
\begin{eqnarray} \label{eq:simpP}
\begin{array}{rcl}
v_0 &=& 0, \\
v_1 &=& -\dfrac{\zeta - \gamma^{\sigma(0)}_{01}}{\sqrt{2\zeta(\zeta-\gamma^{\sigma(0)}_{01})}},\\
v_j &=& \dfrac{\gamma^{\sigma(0)}_{01}}{\sqrt{2\zeta(\zeta-\gamma^{\sigma(0)}_{01})}},\quad j >1,\\
\zeta &=& \pm \rm sgn(\gamma^{\sigma(0)}_{01})\sqrt{\sum_{j>0} \gamma^{2\sigma(0)}_{0j}}.
\end{array}
\end{eqnarray}
We obtain the density-matrix of the cluster $\bar{\gamma}^{\sigma(0)}$ after solving the cluster Hamiltonian $\bar{H}^c_{\rm eff}$ (see \Eq{eq:cluster_Ham}) resulting from the transformation using the previous reflection matrix $\mathbf{R}[\gamma^{\sigma(s=0)}]$. 
For the single impurity case, the cluster 1-RDM is a 2-by-2 matrix, and the density matrix $\gamma^{\sigma(1)}$ of the system is reconstructed using \Eq{eq:trans_inv} to obtain:\\
\begin{eqnarray} \label{eq:gammaz}
\gamma^{\sigma(1)} = \rho^{\sigma(0)} + z^{0}\gamma^{\sigma(0)hd}
\end{eqnarray}explained in \Sec{sec:sces}
Where $ \rho^{\sigma(0)}$ refers to the diagonal part only of $\gamma^{\sigma(0)}$ and $\gamma^{\sigma(0)hd} = \gamma^{\sigma(0)} - \rho^{\sigma(0)}$ to the off-diagonal elements only, and $z^0 = \bar{\gamma}^{\sigma(0)}_{01}/\zeta$.\\
Inserting \Eq{eq:gammaz} in \Eq{eq:simpP}, we obtain the new reflection obtained $\mathbf{R}^{\sigma(1)}[\gamma^{\sigma(1)}]$ that is the same as the previous $\mathbf{R}^{\sigma(0)}[\gamma^{\sigma(0)}]$, in agreement with Appendix C of Ref. \cite{sekaran_householder_2021}. 
Consequently, the self-consistent scheme is achieved at the first step and the kinetic and total interaction energy of the impurity within the full system is exactly the same as the cluster $\bar{H}^c_{\rm eff}$, {\it i.e.} the virial-like relation in \Eq{eq:Virial} is respected.
The profile occupation of natural spin-orbitals $\{\eta_k\}$ is straightforwardly derived\\
\begin{eqnarray}
\begin{array}{rcll}
\eta_k &=& (1+z)/2 &, k\leq k_F,\\
\eta_k &=& (1-z)/2&, k>k_F,
\end{array}
\end{eqnarray}
where $k_F$ corresponds to the natural orbitals at the Fermi level in the non-interacting case. 
This profile is also found in the Fermi-liquid theory, where the renormalization factor $z$ is associated to a change in the effective mass of particles.
The value of the renormalization factor $z=E_K/E^0_K$, with $E_k$ the kinetic energy calculated and $E^0_k=-4/\pi$ the non-interacting kinetic energy at half-filling, depends on the correlation strengh, that goes to 1 in the non-interacting case ($U$=0), to 0 in the atomic limit ($t$=0).\\
In \Fig{fig:single_imp}, we present the renormalization factor $z$ for the single impurity case (this work) compared to the Gutzwiller approximation of the wave-function ~\cite{gutzwiller1965correlation} with respect to the correlation strength $U/t$. The Gutzwiller approximation corresponds to a semi-classical evaluation of the wave-function, where doubly occupied configurations are considered via a statistical average and provides an illustrative example for Fermi-liquid theory. This method, therefore, fails in capturing non-local fluctuations ~\cite{kollar2002exact}, which are predominant for the 1D Hubbard model.
For the single impurity case, the function $z(U/t)$ is convex and strictly decreases with $U/t$, meaning that the occupation profil obtained is always N-representable. At half-filling, we do not anticipate any Mott-Hubbard transition ~\cite{lieb_absence_1968}, with the system retaining its insulating state for all correlation strengths $U/t>0$. The Gutzwiller approximation predicts a Mott-Hubbard transition ~\cite{brinkman1970application} for $U/(U+4t)=0.57$ at half-filling. However, for the single impurity case, the system persists in its metallic state for all correlation strengths $U/t$.
Using \Eq{eq:gammaz}, natural orbitals of the correlated density matrix are preserved, i.e eigenvectors of the correlated 1-RDM obtained are the same as the non-interacting one (Bloch states), and other symmetries such as particle-hole symmetry at half-filling are preserved.
\newpage
\newcommand{\Aa}[0]{Aa}
\end{document}